\documentclass{ws-ijmpa}

\begin{document}

%

\def\nocropmarks{\vskip5pt\phantom{cropmarks}}
\let\trimmarks\nocropmarks      

%


%
%
\setcounter{page}{1}

\title{SPIN ALIGNMENT OF VECTOR MESON\\
         IN HIGH ENERGY REACTIONS}

\author{\footnotesize QING-HUA XU, CHUN-XIU LIU\footnote{
Present address: Institute of High Energy Physics, Beijing 100039, China} 
and ZUO-TANG LIANG}

\address{Physics Department, Shandong University,
Ji'nan, Shandong 250100, China}

\maketitle


\begin{abstract}
The recent data on the polarization of vector meson at LEP
show that the vector mesons favor the helicity zero state.
We calculate the helicity density matrix of vector meson
which contains a polarized fragmenting quark
by adding the spin of the fragmenting quark and that of the
antiquark created in the fragmentation.
The data at LEP imply a significant polarization
for the antiquark in the opposite direction
as that of the fragmenting quark.
We extend the calculations to other reactions and make
predictions for vector mesons
in deeply inelastic lepton-nucleon scatterings and polarized 
$pp$ collisions.
\end{abstract}

\section{Introduction}	

Recently, the OPAL and DELPHI collaborations measured the
polarizations of vector mesons in $e^+e^-$ annihilation at
$Z^0$ pole at LEP.
The data\cite{data1} show clearly that 
the vector mesons in particular those with large momentum fraction
favor the helicity zero state.
Obviously, the study of this subject should provide us 
important information on the hadronization mechanism 
especially the spin effects in fragmentation processes.
Compared with hyperons,
the study of vector meson polarization has two advantages:
First, the decay contribution of heavier hadrons to vector
meson production is much smaller.
Second, the production rate of vector meson is in general 
higer than that of hyperon.
It is thus interesting and important to ask what we can learn from
the data about the polarization of the quarks and antiquarks
created in the fragmentation process.

To answer the above-mentioned question,
in a recent paper\cite{xu01}, we calculated the 
helicity density matrix of vector mesons
by assuming that the spin of vector meson is just
the sum of the spin of the quark and that of the antiquark
which combine into a vector meson.
By comparing the obtained results with the data, 
we found that the LEP data give a strong constraint on the
polarization of the antiquark created in the fragmentation process.
We thus extend our calculations to the deeply
inelastic lepton-nucleon scattering (DIS) process and the high $p_T$
jets in polarized high energy $pp$ collisions.
In this talk, I will make a brief summary of the calculations 
in Ref. 2 and 3, present the main results and make a short discussion. 
\section{Calculation method of helicity density matrix }
To calculate the helicity density matrix of a vector meson
produced in the fragmentation of a polarized quark $q^0_f$,
we divide the produced
vector mesons into the following two groups
and consider them separately:
(a) those which contain the initial quark $q^0_f$'s
or the initial antiquark $\bar{q}^0_f$'s;
(b) those which don't contain the initial quark or antiquark.
The spin density matrix $\rho^V(z)$ for the vector meson $V$
is given by:           
\begin{equation}
\rho^V({\it z})=
\sum_f
\frac {\langle n(z|a,f)\rangle}{\langle n(z)\rangle}
\rho^V(a,f)
+\frac {\langle n(z|b)\rangle}{\langle n(z)\rangle}
\rho^V(b)
.
\label{eq2}
\end{equation}   
Here $\langle n(z|a,f)\rangle$ and $\rho^V(a,f)$
are the average number and spin density matrix 
of vector mesons from (a);
$\langle n(z|b)\rangle$ and $\rho^V(b)$ are those from (b).
$\langle n(z)\rangle$=
$\sum_f \langle n(z|a,f)\rangle$ +
$\langle n(z|b)\rangle$ is the
total number of vector mesons
and $z$$\equiv$$2p_V/\sqrt{s}$, where
$p_V$ is the momentum of vector meson and $\sqrt {s}$ is the total
$e^+e^-$ center of mass energy.
The average numbers $\langle n(z|a,f)\rangle$
and $\langle n(z|b)\rangle$
are determined by the hadronization mechnism and can be
calculated using hadronization models\cite{lund} as implemented by
the Monte-Carlo events generators.   
               
The vector mesons from group (b) are taken as unpolarized
thus $\rho^V(b)$=1/3.
For those from group (a), i.e.,
those which contain the $q^0_f$ (${\bar q}^0_f$)
and a $\bar q$ ($q$) created in the fragmentation process,
$V$=$(q^0_f\bar q)$
or $V$=$({\bar q}^0_f q)$,
the spin density matrix $\rho^V(a,f)$ can be calculated
from the direct product of the spin density matrix
$\rho^{q^0_f}$ for $q^0_f$ (or $\rho^{{\bar q}^0_f}$ 
for ${\bar q}^0_f$) and that $\rho^{\bar{q}}$
for the antiquark $\bar{q}$ (or $\rho^q$ for q).
For explicity, we take $V$=$(q^0_f\bar q)$. 
The spin density matrix of the $q^0_f{\bar q}$ system is
$\rho^{q^0_f{\bar q}}$=$\rho^{q^0_f}$$\otimes$$\rho^{\bar q}$.
Transforming to the coupled basis $|s, s_z\rangle$ 
(where $\vec{s}$=${\vec {s}^q}$+${\vec {s}^{\bar q}}$),
we obtain,                  
\begin{equation}
\rho^V(a,f)=\frac{2}{3+P_fP_z}
\small {
\left (
\begin{array}{ccc}
c_{1f}(1+P_z)                   &
 \frac{c_{1f}}{\sqrt2}(P_x-iP_y)                &
 0                               \\
\frac{c_{1f}}{\sqrt2}(P_x+iP_y) &
 \frac{{1}}{2}(1-P_fP_z)&
 \frac{c_{2f}}{\sqrt2}(P_x-iP_y) \\
0                               &
 \frac{c_{2f}}{\sqrt2}(P_x+iP_y)                &
 c_{2f}(1-P_z)                   \\
\end{array}  \right)
.
}
\end{equation}                
where $c_{1f}$=$(1+P_f)/2$, $c_{2f}$=$(1-P_f)/2$
and $P_f$ is the polarization of the initial quark $q_f^0$;
$\vec{P}(P_x,P_y,P_z$) is
the polarization vector of $\bar{q}$ in the helicity frame of 
$q_f^0$ in which the $z$-axis is taken as the moving direction 
of $q_f^0$.

After transforming the above result to the 
helicity basis of the produced vector meson, 
we obtain the helicity density matrix element $\rho_{00}$, i.e.,
the probability for vector meson in the helicity zero state,
\begin{equation}
\rho_{00}^V(a,f)\approx (1-P_fP_z)/(3+P_fP_z).
\label{eq3}
\end{equation}    
From this result, we can already see that $\rho_{00}^V(z,|a,f)$$>$1/3
only if $P_f$ and $P_z$ have opposite signs.
Both OPAL and DELPHI data\cite{data1} explicitly show that
$\rho^V_{00}$$>$1/3 for all the vector mesons except for $\omega$.
This implies that $P_z$$\neq$0
and has the opposite sign as $P_f$ in these cases. 
\section{Results for $e^+e^-$$\to$$VX$}
We use Eq. (3) to calculate $\rho_{00}^V(a,f)$ in which the
only free parameter is $P_z$.
We first calculate $\rho_{00}$ for $K^{*0}$ as a
function of $z$ since the corresponding data is available\cite{data1}.
To fit the data, the value of $P_z$
has to be quite large. In Fig. 2 we show the results obtained
by taking $P_z$=0.48, which can fit the data reasonably well.

Similiar calculations can certainly be made for other vector
mesons such as $\rho$, $\phi$, $D^{*\pm}$, $B^{*}$ $etc$.
But, for these mesons, only the average values of $\rho_{00}$
in certain $z$ ranges are available, and we integrate the $z$
dependent $\rho_{00}$ over certain $z$ region.
We found out that, for most of the vector mesons,
the resulting $P_z$ can be written as
$P_z$=$-\alpha{}P_f$ with $\alpha{}$ is common for most
of the $q_f^0$'s. The results of
$\alpha{}$=0.51 as determined above from the data
for $K^{*0}$ can be seen in Table 1 of Ref. 2. 
\begin{tabular}{ll}
\begin{minipage}{6cm}
\psfig{file=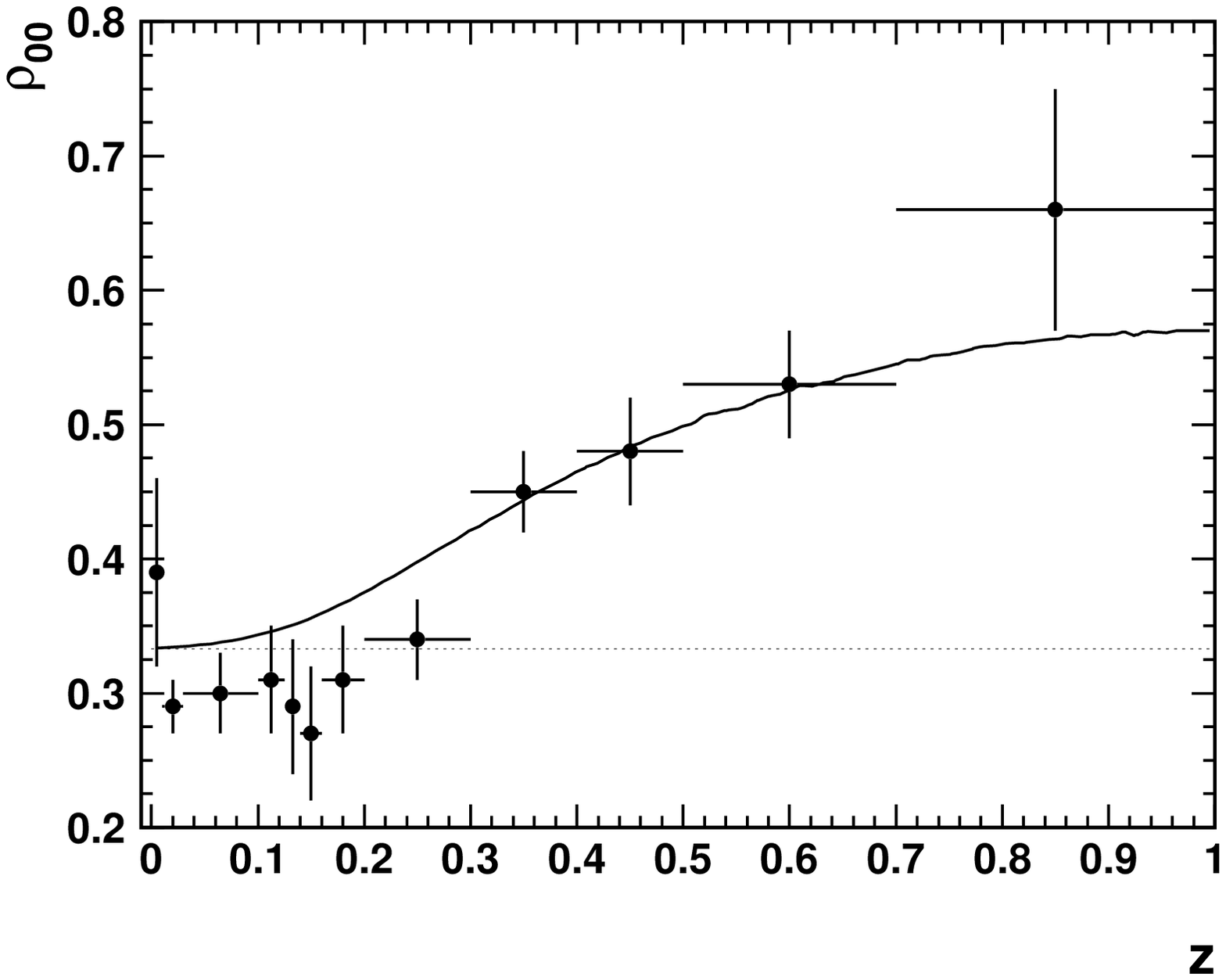,width=5cm} 
\end{minipage}
&
\begin{minipage}{5.5cm}
{\footnotesize Fig.1. Spin density matrix element $\rho_{00}$ of $K^{*0}$
produced in $Z^0$ decay as a function of the momentum fraction $z$.
The solid line is obtained from Eq. (3) by taking $P_z=0.48$;
the dotted line corresponds to the unpolarized case(1/3);
the data are from Ref. 1.}
\end{minipage}
\end{tabular}
\section{Results for $lp$$\to$$l'VX$}
The results we obtained for the polarization of the $\bar q$
in the fragmentation of the longitudinally polarized $q^0_f$
are direct consequences of the LEP data
if we accept that the spin of the produced $V$=$(q^0_f\bar q)$ 
is the sum of the spin of $q^0_f$ and that of $\bar q$.
Since the spin effects in hadronization processes are 
poorly studied yet, it should be interesting to see
whether they are also true in other reactions.
Hence, we extend the calculations to 
polarized DIS processes and $pp$ collisions.

Now, we present the calculations in $lp \to l'VX$. 
Here we do the calculations in the
center of mass frame of the hadronic system, 
and replace the momentum fraction $z$ used 
in $e^+e^-$$\to$$VX$ by $x_F$$\equiv$$2p_z^*/W$, 
where $z-$axis is taken as the moving
direction of the intermediate boson.
Using the relation in Eq.(3) and Eq.(1), we obtain the 
$\rho^V_{00}$ for DIS process,
\begin{equation}
\label{eq5}
\rho_{00}^V(x_F)=\sum_f
\frac{1+\alpha P_f^2}{3-\alpha P_f^2}
\frac {\langle n(x_F|a,f)\rangle}{\langle n(x_F)\rangle}
+ \frac{1}{3}
\frac {\langle n(x_F|b)\rangle}{\langle n(x_F)\rangle}
.
\end{equation}    
The $P_f$ now is the longitudinal polarization of the outgoing
struck quark.
We calculate $\rho_{00}^V$ 
in the current fragmentation region ($x_F$$>$0) of $\mu^-p$$\to$$\mu^-VX$
at $E_\mu$=500 GeV and 
the results for different combinations of 
beam polarization $P^{l}$ and target polarization $P^{N}$
are shown in Fig. 2 where $\alpha$=0.5 is used.
We see that when the lepton beam is polarized, the $\rho_{00}^V$
are much larger than $1/3$ for large $x_F$ region.   

For the neutrino DIS process
$\nu_{\mu} N$$\to$$ \mu^- VX$,
$P_f$=-1 (or 1) for the outgoing struck quark (antiquark),
which reaches the maxmum.
The results of $\rho_{00}^V$ for different mesons
can be found in Ref. 3.
The spin alignments for $K^{*+}$, $\rho^{\pm}$ and $\rho^{0}$
increase to about 0.6 for large $x_F$ region.  

\begin{tabular}{ll}
\begin{minipage}{6cm}
\psfig{file=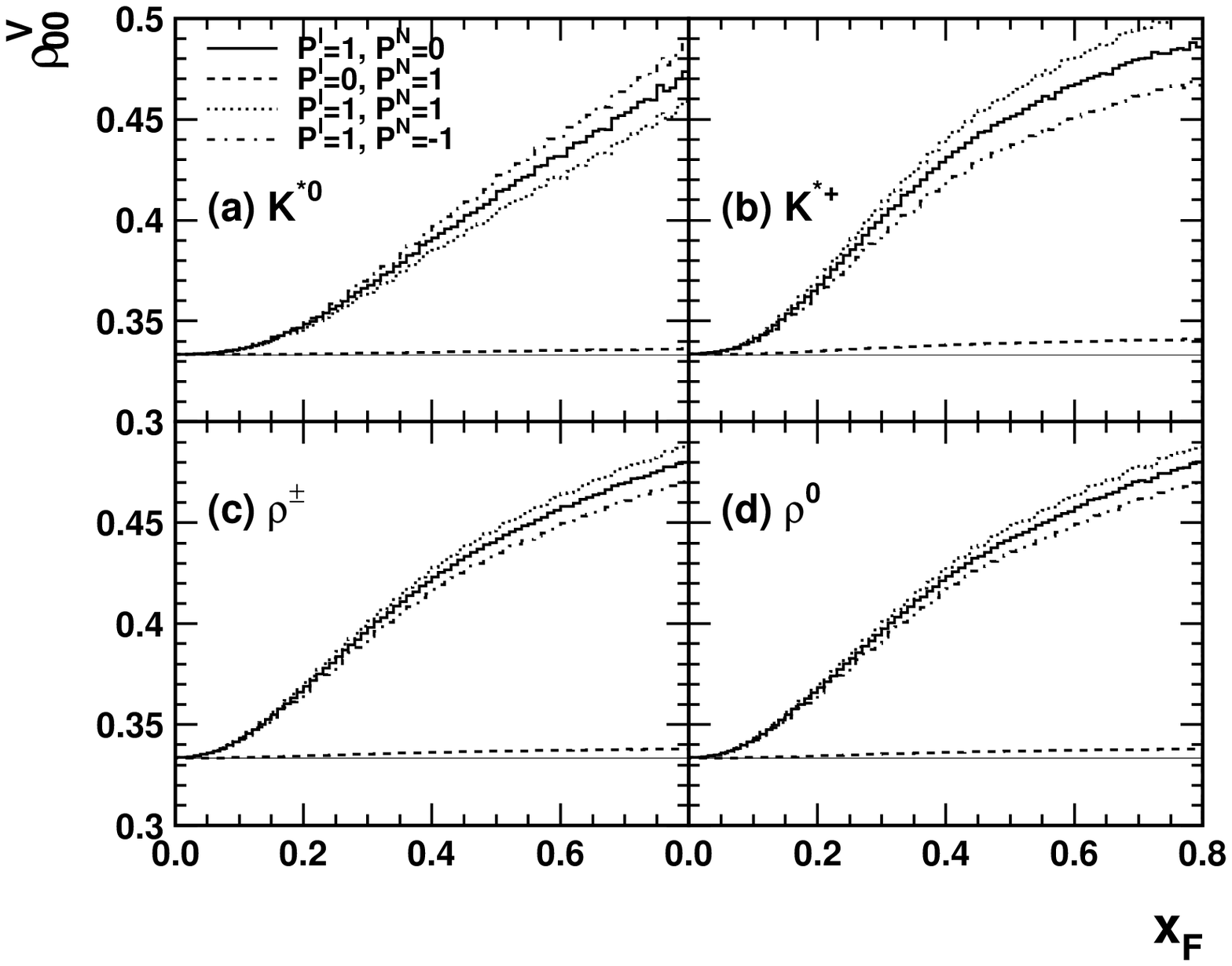,width=4.4cm}
\end{minipage}
&
\begin{minipage}{5cm}
{\footnotesize
Fig.2. Spin alignments of different vector mesons in the current
fragmentation region of $\mu^- p$$ \to$$ \mu^- VX$ at $E_\mu$=500 GeV.
The straight line corresponds to $\rho_{00}$=1/3.}
\end{minipage}
\end{tabular}
\section{Results for $pp$ collisions}
In high energy $pp$ collisions, such as at RHIC,
the vector mesons in high $p_T$ jets can be considered as
the fragmentation products of the outgoing quarks in the 
hard scattering subprocess. 
When the beam or/and target proton is longitudinally polarized, 
the polarization of the incoming partons can be known from the
polarized distribution functions\cite{GRSV2000} and 
the polarization of the outgoing partons  
can be calculated by perturbative QCD.
Our results of spin alignments for vector mesons 
with $p_T$$>$13 GeV at $\sqrt s$=500 GeV
as a function of rapidity $\eta$
in beam polarized $pp$ collisions are shown in Fig.3
where the same $\alpha$ is used as in previous cases.
\begin{tabular}{ll}
\begin{minipage}{6cm}  
\psfig{file=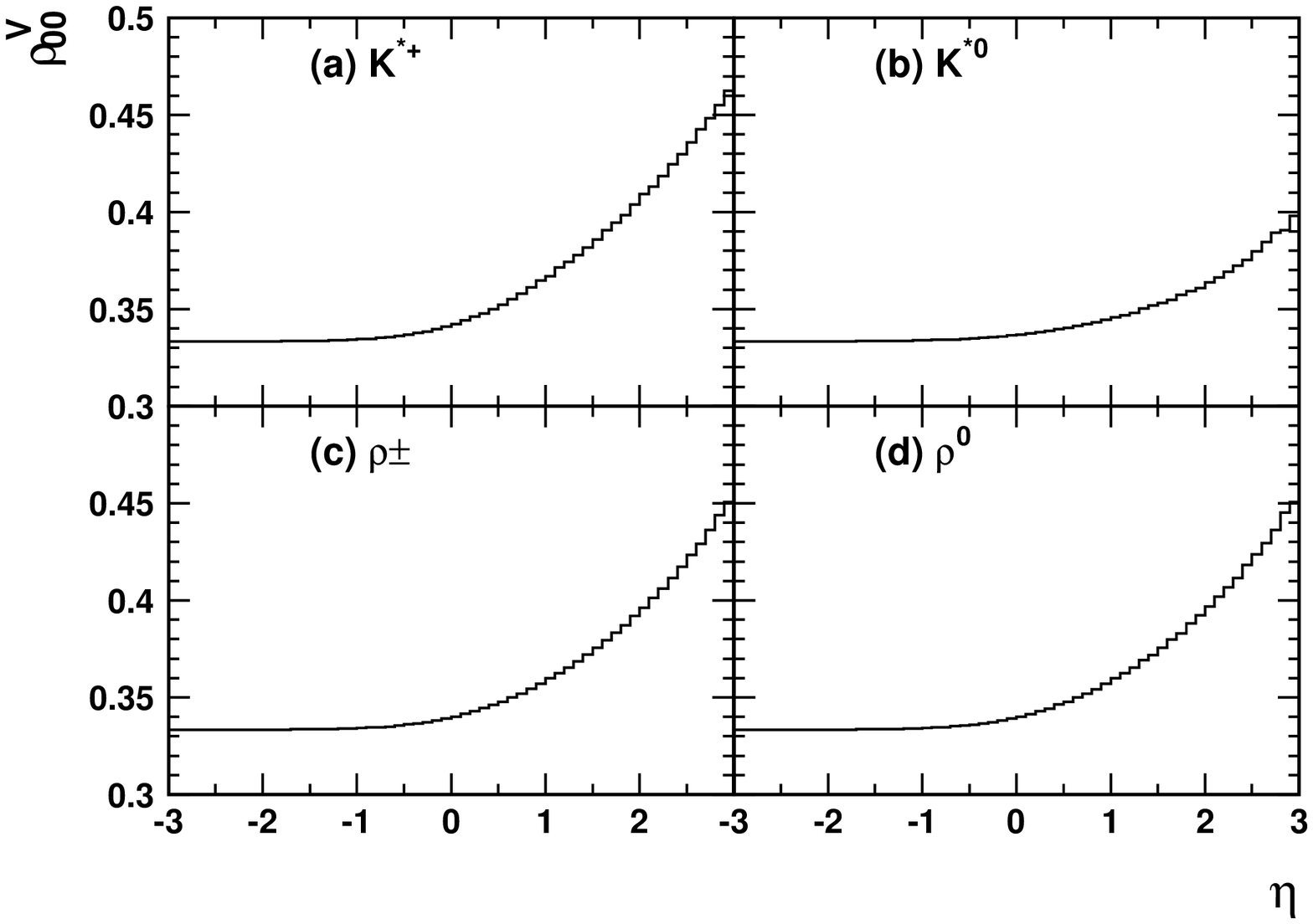,width=4.8cm}
\end{minipage}
&
\begin{minipage}{4.5cm}
{\footnotesize        
Fig.3. Spin alignments of different vector mesons of $p_T$$>$13 GeV 
in $\vec p p$$\to$$VX$ at
$\sqrt s$=500 GeV. }
\end{minipage}
\end{tabular} 

\section*{Acknowledgements}
This work was supported in part by the National Science Foundation
of China (NSFC) and the Education Ministry of China.

\end{document}